\documentstyle[sprocl,epsfig]{article}

\def\NPB{{\em Nucl. Phys.} B}
\def\PLB{{\em Phys. Lett.}  B}

\def\PRD{{\em Phys. Rev.} D}
\def\ZPC{{\em Z. Phys.} C}

\def\be{\begin{equation}}
\def\ee{\end{equation}}
\def\bea{\begin{eqnarray}}
\def\eea{\end{eqnarray}}

\begin{document}

\title{HEAVY QUARKS AND STRUCTURE FUNCTIONS}

\author{R. S. THORNE}

\address{Department of Physics, Theoretical Physics, 1 Keble Road,\\
Oxford OX1 3NP, England\\E-mail: thorne@v2.rl.ac.uk}


\maketitle\abstracts{I present a method for calculating heavy
quark production which extrapolates smoothly from the well understood limits 
of the fixed-flavour number scheme for $Q^2\sim m_H^2$ to the zero-mass 
scheme at $Q^2/m_H^2 \to \infty$. For all $Q^2> m_H^2$ the evolution of the 
heavy quark distribution is precisely as in the 
massless $\overline{\hbox{\rm MS}}$
scheme. The method is simple to implement and compares well with data.}

\section{Introduction}

Until recently the treatment of heavy flavours in global analyses of 
structure functions has been rather 
naive. Usually the charm and bottom quarks were 
regarded as infinitely massive below $Q^2=m_H^2$, and massless above this 
threshold. This procedure, known as the zero mass variable flavour number
scheme (ZM-VFNS), gives the correct results for $Q^2/m_H^2 \to
\infty$, but has errors of ${\cal O}(m_H^2/Q^2)$ and is very inaccurate 
near threshold. Alternatively, all heavy flavour final states may be thought
of as being produced by the hard scatter between the electroweak boson 
and a light parton, i.e. the number of active flavours, $n_f$ is 3 and the
heavy flavour cross-section is generated (mainly) by photon-gluon fusion. 
This is called the fixed flavour number scheme (FFNS), and incorporates the 
correct threshold behaviour, i.e. a smooth kinematic threshold of 
$W^2=Q^2(x^{-1}-1)=4m_H^2$, automatically. This method is certainly appropriate
provided $Q^2$ is not too large.  However, it does not 
sum potentially large logarithms in $Q^2/m_H^2$, and for this reason (and also 
because of the related lack of a heavy flavour parton distribution)
is unsuitable for $Q^2\gg m_H^2$.  

Recently there have been measurements of the charm structure function at 
HERA \cite{hone,zeus}, complementing
the previous measurements by EMC \cite{emc}. Also, the 
charm contribution to the total $F_2$ at small $x$ at HERA can be $> 20\%$,
and the NMC data, which contains only $5-10\%$ charm, but has very small 
errors, is surprisingly sensitive to the treatment of charm. 
Hence, a global 
analysis of the most up to date structure function data must 
necessarily include a description of heavy flavours 
which is satisfactory over a wide range of $x$ and $Q^2$.

\section{Theoretical Procedure}

We desire an approach which extrapolates smoothly from the FFNS at low $Q^2$ 
to the ZM-VFNS at high $Q^2$, maintaining the correct ordering in both 
schemes. In order to do this we first examine the relationship between the 
parton densities in these two schemes (putting $\mu^2=Q^2$ for simplicity). 
One can show that the 4 flavour 
partons in the ZM-VFNS distribution are related to those in the 3 flavour
FFNS by the equation
\begin{equation}
f^{4}_k(Q^2,m_c^2)=A_{ka}(Q^2/m_c^2)\otimes f^{3}_a(Q^2),
\label{eq:one}
\end{equation}
where the matrix elements $A_{ka}(Q^2/m_c^2)$ are calculable in 
perturbation theory, and contain logs in $Q^2/m_c^2$.\footnote{At NLO (but
not beyond) the charm distribution starts from zero at $Q^2=m_c^2$ and all
other partons are the same in both schemes at this scale.}  Hence, 
the charm density is totally defined in terms of the light parton densities. 
Due to the formal equality of the two schemes for $Q^2/m_c^2\to\infty$, 
in this limit we find a reciprocal relationship for the coefficient
functions
\begin{equation}
C^{FF}_{2,ca}(Q^2/m_c^2)=C^{4}_{2,ck}\otimes A_{ka}(Q^2/m_c^2).
\label{eq:two}
\end{equation}
So in this limit all the logs in $C^{FF}_{2,ca}(Q^2/m_c^2)$ can be factored 
into the matrix elements $A_{ka}(Q^2/m_c^2)$, and absorbed into the definition 
of the 4-flavour parton densities where they are summed automatically by
evolution \cite{buza}.  
 
For the scales $Q^2<m_c^2$ we assume the validity of the FFNS approach. 
For $Q^2>m_c^2$ we extend the approach outlined above 
to define a variable flavour number scheme (VFNS). We make precisely the 
same definition of the 4-flavour parton 
densities as above (thus ensuring that they
evolve as massless partons), but in order to incorporate 
threshold effects correctly, i.e. 
guarantee exact
equivalence to the FFNS, we have mass dependent coefficient functions. In 
particular we demand that for all $Q^2>m_c^2$ our coefficient functions 
satisfy 
\begin{equation}
C^{FF}_{2,ca}(Q^2/m_c^2)=C^{VF}_{2,ck}(Q^2/m_c^2)\otimes A_{ka}(Q^2/m_c^2).
\label{eq:three}
\end{equation}

Unfortunately, because there are more degrees of 
freedom in our coefficient functions than there are equations determining 
them, they are not defined uniquely. For example at lowest order in $\alpha_s$
we have the equation \cite{acot}
\begin{equation}
C^{FF,1}_{2,cg}(Q^2/m_c^2)=C^{VF,0}_{2,cc}(Q^2/m_c^2)\otimes 
A_{cg}(Q^2/m_c^2)+C^{VF,1}_{2,cg}(Q^2/m_c^2).
\label{eq:four}
\end{equation}
Since in the correct ordering of $F_{2,c}(x,Q^2)$ only the 
zeroth order coefficient functions appear at LO, the first 
order coefficient functions appearing at NLO, the way in which we choose to 
solve (\ref{eq:four}), and higher order equations, affects our order-by-order
expressions for $F_{2,c}(x,Q^2)$. We remove this ambiguity by 
imposing the physical requirement that at the transition point $Q^2=m_c^2$, 
where we go from a 3-flavour to 4-flavour description, 
$\partial F_{2,c}(x,Q^2)/\partial \ln Q^2$ 
is continuous. Imposing this to all orders (in the gluon sector)
uniquely determines all coefficient functions and defines 
our VFNS.\footnote{We note that both our coefficient functions and our 
definition of ordering is different to the alternative,
currently LO, VFNS \cite{acot}.} 
Imposing this constraint, the LO coefficient function is given by the 
relatively simple expression
\begin{equation}
\partial C^{FF,1}_{2,cg}(Q^2/m_c^2)/\partial \ln Q^2=
C^{VF,0}_{2,cc}(Q^2/m_c^2)\otimes P^0_{qg},
\label{eq:five}
\end{equation}
where $P^0_{qg}$ is the LO quark-gluon splitting function. This then 
automatically defines 
$C^{VF,1}_{2,cg}(Q^2/m_c^2)$ via (\ref{eq:four}). $C^{VF,1}_{2,cc}(Q^2/m_c^2)$
is more complicated, but numerically unimportant. The treatment for bottom 
quarks is essentially identical, and the procedure can easily be 
generalized to other processes. Further 
details can be found in \cite{rt}.

\begin{figure}
\centerline{\epsfig{figure=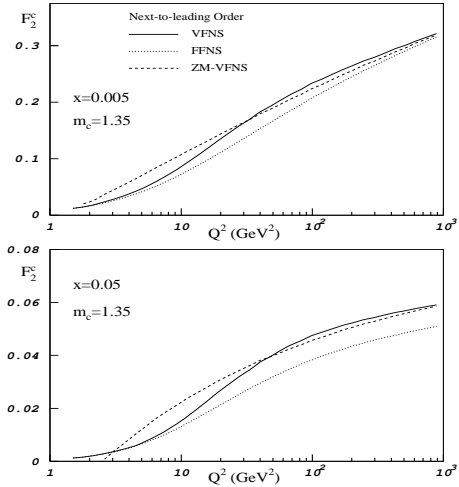,height=3in,width=2.5in}}
\vspace{-0.5in}
\caption{Comparison of $F_{2,c}(x,Q^2)$ using different schemes.}
\label{fig:charmfig1}
\end{figure}

\section*{Results}
\begin{figure}
\centerline{\epsfig{figure=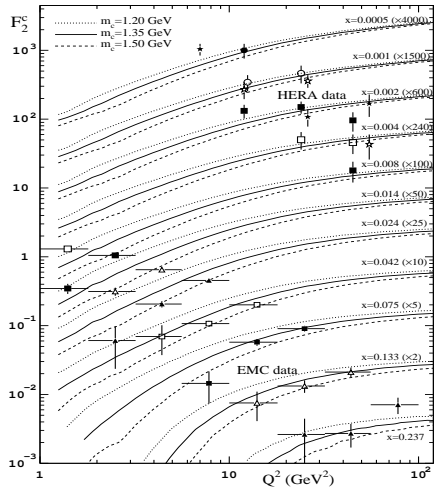,height=3in,width=2.5in}}
\vspace{-0.5in}
\caption{Prediction for $F_{2,c}(x,Q^2)$ compared to data.}
\label{fig:charmfig2}
\end{figure}

Fig. \ref{fig:charmfig1} shows the results of the NLO calculation of 
$F_{2,c}(x,Q^2)$, illustrating how it smoothly extrapolates from the FFNS
result at $Q^2 \sim m_c^2$ to the ZM-VFNS at high $Q^2$. We have also 
performed global fits using different prescriptions for heavy flavours and 
find that the best quality fit is indeed obtained using our VFNS 
approach \cite{rt}.
Fig. \ref{fig:charmfig2} shows the prediction for $F_{2,c}(x,Q^2)$ 
obtained from the parton distributions, resulting from our best fit to the
total $F_2(x,Q^2)$ (using $m_c=1.35 {\hbox{\rm GeV}}$), for a variety of 
values of $m_c$. Again one can see the smoothness and the good 
agreement with the full range of available data. 

Thus, our VFNS for heavy quark production is on a firm theoretical footing, 
it is quite simple to implement, clearly 
has the behaviour we would expect, and matches well to data. Hence we   
encourage its use in analyses of structure functions.

\end{document}